%% file: charm2015_vogt.tex
\documentclass[12pt]{article}
\usepackage{graphicx}
\usepackage{subfig}

\newcommand\pubnumber{LLNL--CONF--675922}
\newcommand\pubdate{\today}

\def\llnl{Nuclear and Chemical Sciences Division\\ Lawrence Livermore
National Laboratory, Livermore,CA 94551, USA}
\def\ucd{Physics Department, University of California at Davis, Davis, CA 
95616, USA}
\def\support{\footnote{This work was performed under the auspices of the U.S.\
Department of Energy by Lawrence Livermore National Laboratory under
Contract DE-AC52-07NA27344 and supported by the U.S. Department of Energy, 
Office of Science, Office of Nuclear Physics (Nuclear Theory) under contract 
number DE-SC-0004014.}}

\def\Title#1{\begin{center} {\Large #1 } \end{center}}
\def\Author#1{\begin{center}{ \sc #1} \end{center}}
\def\Address#1{\begin{center}{ \it #1} \end{center}}

\newcommand\pubblock{\rightline{\begin{tabular}{l} \pubnumber\\
         \pubdate  \end{tabular}}}
\newenvironment{Abstract}{\begin{quotation}  }{\end{quotation}}
\newenvironment{Presented}{\begin{quotation} \begin{center} 
             PRESENTED AT\end{center}\bigskip 
      \begin{center}\begin{large}}{\end{large}\end{center} \end{quotation}}

\input econfmacros.tex
\begin{document}
\begin{titlepage}
\pubblock

\vfill
\Title{Cold Nuclear Matter Effects on Open and Hidden Heavy Flavor 
Production at the LHC}
\vfill
\Author{R. Vogt\support}
% put in address(es) defined above
\Address{\llnl \ucd}
\vfill
\begin{Abstract}
We discuss a number of cold nuclear matter effects that can modify open
heavy flavor and quarkonium production in proton-nucleus collisions and could 
thus also affect their production in nucleus-nucleus collisions, in addition
to hot quark-gluon plasma production.  We show some results for $p+$Pb
collisions at $\sqrt{s} = 5$ TeV at the LHC.
\end{Abstract}
\vfill
\begin{Presented}
The 7th International Workshop on Charm Physics (CHARM 2015)\\
Detroit, MI, 18-22 May, 2015
\end{Presented}
\vfill
\end{titlepage}
\def\thefootnote{\fnsymbol{footnote}}
\setcounter{footnote}{0}
%

%%%%%%%%%%%%%%%%%%%%%%%%%%%%%%%%%%
%\section{Introduction}

High energy heavy-ion collisions, especially those at collider energies at
RHIC and the LHC, study the low baryon density and high temperature region of
the phase diagram of nuclear matter.  These experiments are designed to search
for phase transitions from normal matter to a deconfined quark-gluon plasma.
There are other mechanisms that can mimic some of the effects of
the deconfinement phase transition that do not involve quark-gluon plasma
production.  The way to separate these 'cold nuclear matter' effects from
those of the deconfined media is to study other, smaller collision systems.
Thus a complete study of quark-gluon plasma production and evolution includes
proton-proton and proton-nucleus collisions at, if not the same energy, as
close as possible.  At RHIC, where the beams are independent, $pp$, d+Au and
Au+Au have been studied at the same nucleon-nucleon center of mass energy, 
$\sqrt{s_{_{NN}}} = 200$ GeV, as well as other, lower energies.  The LHC
collider has studied  is set up so that the beams are not independent, thus the
intermediate $p+$Pb collisions run in 2013 involved a 4 TeV proton beam on
a $4(Z_{\rm Pb}/A_{\rm Pb}) = 1.58$ TeV lead beam.  The resulting 
$\sqrt{s_{_{NN}}} = 5.02$ TeV $p+$Pb system is close in energy to the upcoming
5.1 TeV Pb+Pb collisions in LHC Run 2.  However, the center of rapidity in
these collisions is shifted backward by $\Delta y = 0.46$ units relative to
the symmetric $pp$ and Pb+Pb collisions run in Run 1 at $\sqrt{s} = 2.76$, 7
and 8 TeV ($pp$) and $\sqrt{s_{_{NN}}} = 2.76$ TeV with forward $y$ defined as
the direction of the proton beam.  In addition, the LHCb
and ALICE detectors which have muon spectrometers on only one side of the
detector require two modes of running, one in which the proton beam travels
toward forward rapidity and one in which it travels toward negative rapidity.

Deviations from the $pp$ baseline in $p+$Pb and Pb+Pb collisions have been 
typically been quantified by ratios of observables in the heavy systems to the
light system, such as the nuclear modification factor,
\begin{eqnarray}
R_{p{\rm Pb}}(y,p_T) = \frac{d\sigma_{p{\rm Pb}}(y,p_T)/dy dp_T}{T_{p{\rm Pb}}
d\sigma_{pp}(y,p_T)/dy dp_T} \, \, .
\label{eqn:rpa}
\end{eqnarray}
So far, no $pp$ collisions have been run at the same energy.  Thus the $p+$Pb
data have been compared to an interpolation of the $pp$ results.  An alternate
way of separating the cold matter effects in these collisions is the use of
the forward-backward ratio,
\begin{eqnarray}
R_{FB}(y,p_T) = 
\frac{d\sigma_{p{\rm Pb}}(y>0,p_T)/dy dp_T}{d\sigma_{p{\rm Pb}}(y<0,p_T)/dy dp_T
} 
\, \,  \, ,
\label{eqn:rfb}
\end{eqnarray}
which requires no interpolation and cancels systematic ratios in the same
system.  

Interpreting the results of these collisions requires an understanding
of the cold matter effects which are important in their own right.  The focus
has been on hard probes (with a large mass or transverse momentum scale)
of these collisions which, although produced early
in the collision history, can be influenced by the bulk which is assumed to
reach thermal equilibrium so that its evolution can be described by 
hydrodynamics.  A special advantage of hard probes is that their production,
in $pp$ collisions, and with suitable modifications, as discussed in the
remainder of these proceedings, can be described by hard scattering in 
perturbative QCD.  Already nearly thirty years ago, $J/\psi$ production was
predicted to be suppressed by deconfinement in nucleus-nucleus collisions
\cite{MatsuiSatz}.  It soon became clear that suppression of a sort was already
present in fixed-target proton-nucleus collisions and whatever was causing this
suppression may also be causing at least some of the suppression in heavy-ion
collisions.  Since the $J/\psi$ is more easily detected by a peak in the
dilepton continuum at 3.097 GeV, it has been studied longer than open charm
hadrons in the same systems.  The best studied charm mesons are 
$D^0/\overline D^0$ since they primarily decay into two light hadrons, 
$K^\pm \pi^\mp$, and can thus be reconstructed, even in the high multiplicity
environment of high-energy heavy-ion collisions.  However, the large energies
of the LHC allow higher statistics measurements of both open and hidden
charm, permitting detailed studies of other charm mesons and an extension of
the charm studies to heavier bottom.   

In this brief overview, we will describe some of the most studied cold nuclear
matter effects.  These include: isospin (the difference between neutrons and 
protons as projectiles); modifications of the parton distributions in the
nuclear medium (shadowing); energy loss as the produced $c \overline c$ pair
travels through the medium (which can manifest itself as a shift in
the longitudinal momentum or as a broadening of the transverse momentum
distribution); intrinsic charm in the proton; and, more specific
to quarkonium final states absorption by nucleons and breakup by interactions
with produced, comoving particles.  We will briefly touch upon all of these
except for intrinsic charm, covered in a recent review \cite{IC_review}.

Isospin affects all perturbative QCD calculations because a nucleus is not
made up of all protons or all neutrons.  It is negligible for $J/\psi$ in most
regions of phase space because $J/\psi$ production is dominated by $gg$ 
processes except in the most forward regions.  However, it is very important
for Drell-Yan, direct photon and gauge boson production.

\begin{figure}[htb]
%\centering
\begin{center}
\includegraphics[width=0.425\textwidth]{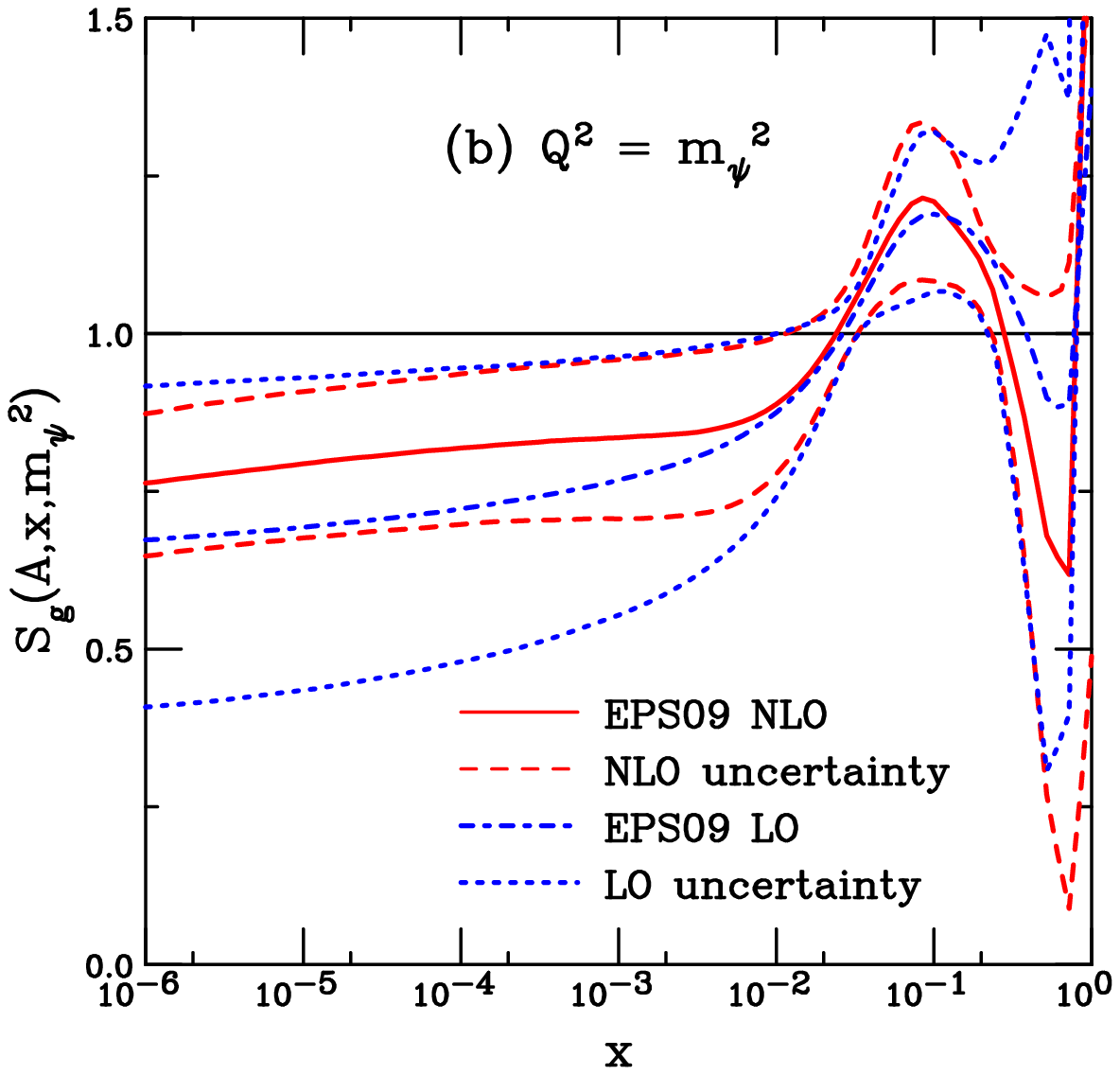} 
%\\
%\includegraphics[width=0.425\textwidth]{eps09glu_nlovslo_ups.ps}
\includegraphics[width=0.425\textwidth]{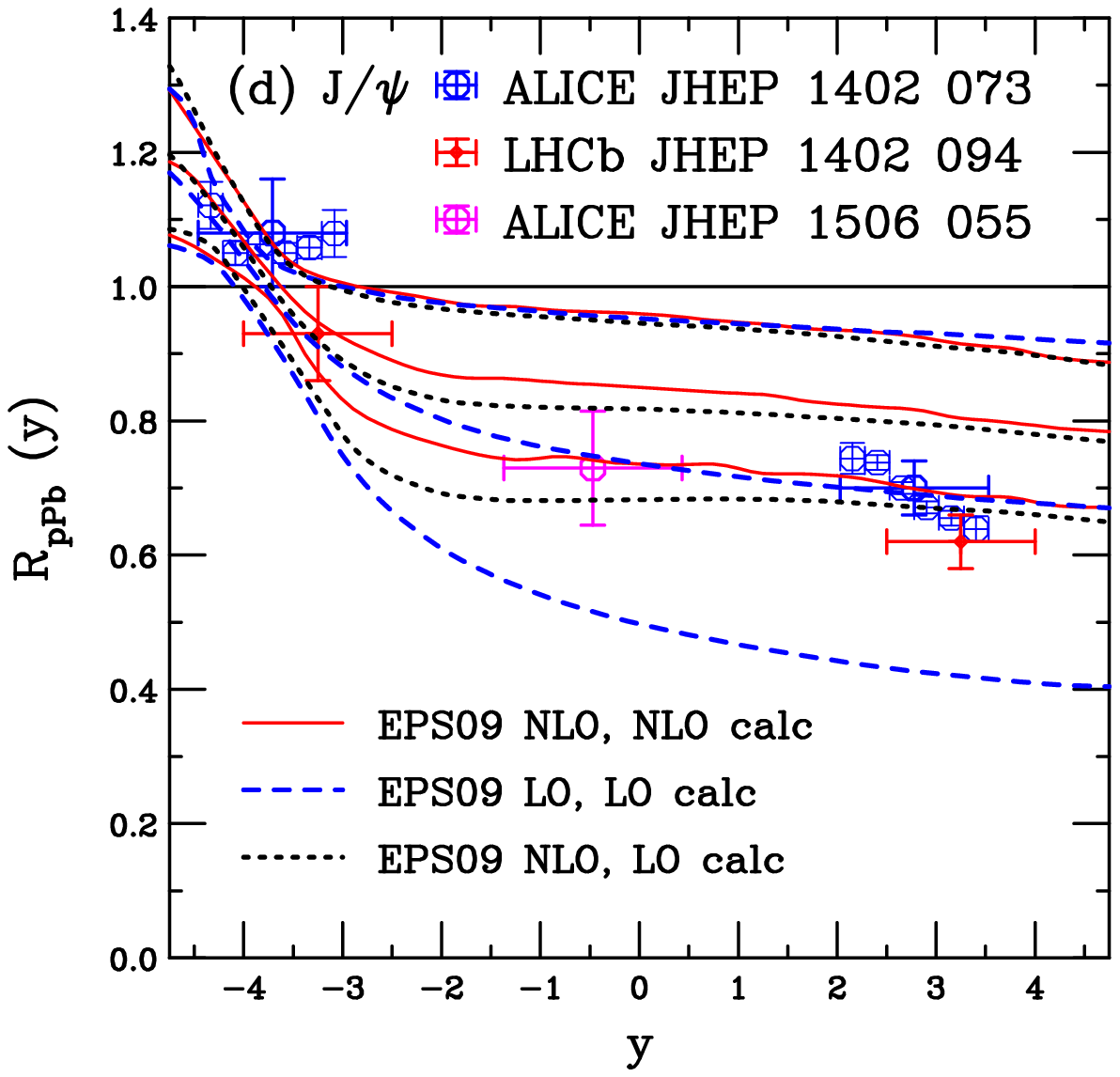}
\end{center}
\caption{(Color online)
The EPS09 LO (blue bands) and NLO (red bands) nuclear gluon
distributions for a lead nucleus, $A = 208$, are presented as a function of
%momentum fraction $x$ for the scale values: (a) $Q^2 = Q_0^2 = 1.69$
%GeV$^2$; (b) $Q^2 = m_{J/\psi}^2$; and (c) $Q^2 = m_\Upsilon^2$.  (Note that the 
%scale values for $m_{J/\psi}$ and $m_\Upsilon$ are the values for the central 
$x$ for $Q^2 = m_{J/\psi}^2$ (a).  Panel (b) shows calculations of the
$J/\psi$ nuclear modification factor in $p+$Pb collisions at 
$\sqrt{s_{_{NN}}} = 5.02$ TeV compared to results with EPS09
NLO in a NLO calculation (red); EPS09 LO in a fully LO calculation (blue); and
EPS09 NLO and a fully LO calculation (black) compared to the ALICE 
\protect\cite{ALICEpPbpsi,ALICEpPbpsi_pT} and LHCb \protect\cite{LHCbpPbpsi}
data.  (Note that the 
scale values for $m_{J/\psi}$ are the values for the central 
renormalization and factorization scales from the open charm fits
\protect\cite{NVF}.) 
}
\label{fig:nPDF}
\end{figure}

The modification of the parton distributions in nuclei, often
referred to as shadowing, also affects production of all hard probes.  
Since nuclear deep inelastic scattering 
experiments probe only charged parton densities, the nature and
magnitude of the effect on the nuclear gluon density is not well known.  
The effects of shadowing in nuclei are parameterized by various groups using
global fitting methods similar to those used to evaluate the parton densities 
in the proton.  Currently, LO and NLO sets are available, evolving quarks, 
antiquarks and gluons separately with $Q^2$.  For example, the EPS09
release \cite{Eskola:2009uj} includes 31 error sets.
The EPS09 uncertainty band is obtained by calculating the deviations
from the central value for the 15 parameter variations on either side of the 
central set and adding them in quadrature.  The results for the EPS09 LO and
NLO gluon modifications at the $J/\psi$ production scale are shown in
Fig.~\ref{fig:nPDF}(a).  There are considerable differences between the LO and
NLO extractions which, however, do not need to manifest themselves as different
observables.  Indeed, in a consistent global analysis, the LO and NLO shadowing
ratios for $J/\psi$ should agree.  This is not quite true for the EPS09 set,
as shown in Fig.~\ref{fig:nPDF}(b) for $R_{p{\rm Pb}}(y)$ at 5.02 TeV.    
The differences between the shadowing ratios calculated for
$J/\psi$ production with the LO set at leading order and the NLO set at 
next-to-leading order demonstrate this fact.  This difference is not an artifact
of the more complex NLO process since applying the NLO set to the LO calculation
gives a result quite similar to the consistent NLO calculation.  The variation
in the same can be attributed to the $2 \rightarrow 3$ processes dominating
$c \overline c$ production at high energies rather than the $2 \rightarrow 2$
in the LO color evaporation model.  The scales are also somewhat different
in the two calculations since the NLO calculation in integrated over the full
$p_T$ range.  For more details, see Ref.~\cite{RV_ppb}.

%%%%%%%%%%%%%%%%%%%%%%%%%%%%%%%%%%%%%%%%%%%%%%%%%%%%%%%%%%%%%%%%%%%%%%%%%%%
%%%%%%%%%%%%%%%%%%%%%%%%%%%%%%%%%%%%%%%%%%%%%%%%%%%%%%%%%%%%%%%%%%%%%%%%%

The dependence of nuclear shadowing on collision centrality (central collisions
have almost full overlap of the two nuclei while more peripheral collisions 
have smaller overlap) has recently 
been studied in heavy-ion collisions.  The PHENIX
d+Au $J/\psi$ data suggested a stronger dependence than expected from a linear
dependence on the path length of the probe through the nucleus 
\cite{Adare:2010fn}.  We checked whether some power $n$ of the nuclear 
thickness function, $T_A(r_T)$, 
at impact parameter $r_T$,
\cite{McGlinchey:2012bp}
\begin{eqnarray}
R_g^{T_A^n}(r_T) = 1 - (1 - R_g(x,Q^2)\,)\bigg( \frac{T_A^n(r_T)}{a(n)} \bigg) 
\, \, .
\label{eqn:modshad_power}
\end{eqnarray}
could describe the data.
Here $R_g(x,Q^2)$ is the EPS09 NLO central set and $a(n)$ is a 
normalization adjusted to give the average
EPS09 modification when integrated over $r_T$.
The PHENIX data preferred $n > 10$ in 
Eq.~(\ref{eqn:modshad_power}).  Since this result is physically unsatisfying, 
also We found that assuming shadowing is concentrated in the core of the nucleus
provides a good description of the data.  
The formulation for gluon shadowing in the core of the
nucleus is \cite{McGlinchey:2012bp}
\begin{eqnarray}
R_g^{\rm core}(r_T) = 
1 - \bigg( \frac{1-R_g(x,Q^2)}{a(R,d)(1 + \exp((r_{_{T}} - R)/d))} \bigg) \,\, .
\label{eqn:modshad_sharp}
\end{eqnarray} 
Here the normalization $a(R,d)$ is again adjusted to give the appropriate
average EPS09 modification.  The core region is found to be defined by a 
radius of $R\sim 2.4$~fm with a relatively sharp cutoff, a width of
$d\sim 0.12$~fm.  Work on a similar calculation for $J/\psi$ production
as a function of rapidity is in progress for the LHC $p+$Pb data.
More data on more final states are needed to test these results.
%Figure~\ref{rtmods} shows the behavior of the EPS09 central gluon ratio,
%$R_g(r_T)$, from fitting the parameters of Eqs.~(\ref{eqn:modshad_power})
%and (\ref{eqn:modshad_sharp}) to the PHENIX data.  
%The result is shown for all the PHENIX rapidity
%bins.  The result for the
%recent EPS09s  spatially-dependent 
%modifications which retain up to quartic powers in
%the expansion of the centrality dependence as a function of path length for
%$A$-independent coefficients \cite{Helenius:2012wd}
%is also shown.  Its dependence is much weaker than either of the fits to
%Eqs.~(\ref{eqn:modshad_power})
%and (\ref{eqn:modshad_sharp}),
The result for the EPS09s spatially-dependent modifications, which retain up 
to quartic powers in the path length of the $A$-independent coefficients 
\cite{Helenius:2012wd}, has a much weaker centrality dependence than either of 
the fits to Eqs.~(\ref{eqn:modshad_power}) and (\ref{eqn:modshad_sharp}),

Initial-state energy 
loss has been studied in Drell-Yan production and found to be small.
However, there is an inherent ambiguity when applying
initial-state energy loss to Drell-Yan production since most groups
parameterizing the nuclear parton densities include these same Drell-Yan data 
to extract the strength of shadowing on the
antiquark densities \cite{Eskola:2009uj}.  Also, by forcing the loss to be 
large enough to explain the high $x_F$ behavior of $J/\psi$ production in 
fixed-target interactions \cite{Leitch:1999ea} violates the upper bound on 
energy loss established by small angle forward gluon emission 
\cite{Brodsky:1992nq}.  More recently, it has been proposed that
cold matter energy loss should be treated
as a final-state effect \cite{Arleo:2012rs}.  The final-state $J/\psi$ 
energy loss in $pA$ collisions is currently implemented as a
probability distribution dependent on the energy loss
parameter. The effect modifies the $x_F$ and $p_T$ distributions in a rather
crude fashion since the quarkonium distribution in $pp$ collisions is 
parameterized as a convolution of factorized power laws, $ \propto (1-x)^n
(p_0^2/(p_0^2 + p_T^2))^m$, rather than using a quarkonium production model 
\cite{Arleo:2012rs,Arleo:2013zua}.  

%%%%%%%%%%%%%%%%%%%%%%%%%%%%%%%%%%%%%%%%%%%%%%%%%%%%%%%%%%%%%%%%%%%%%%%%%
\begin{figure}[htb]
\centering
\includegraphics[width=0.425\textwidth]{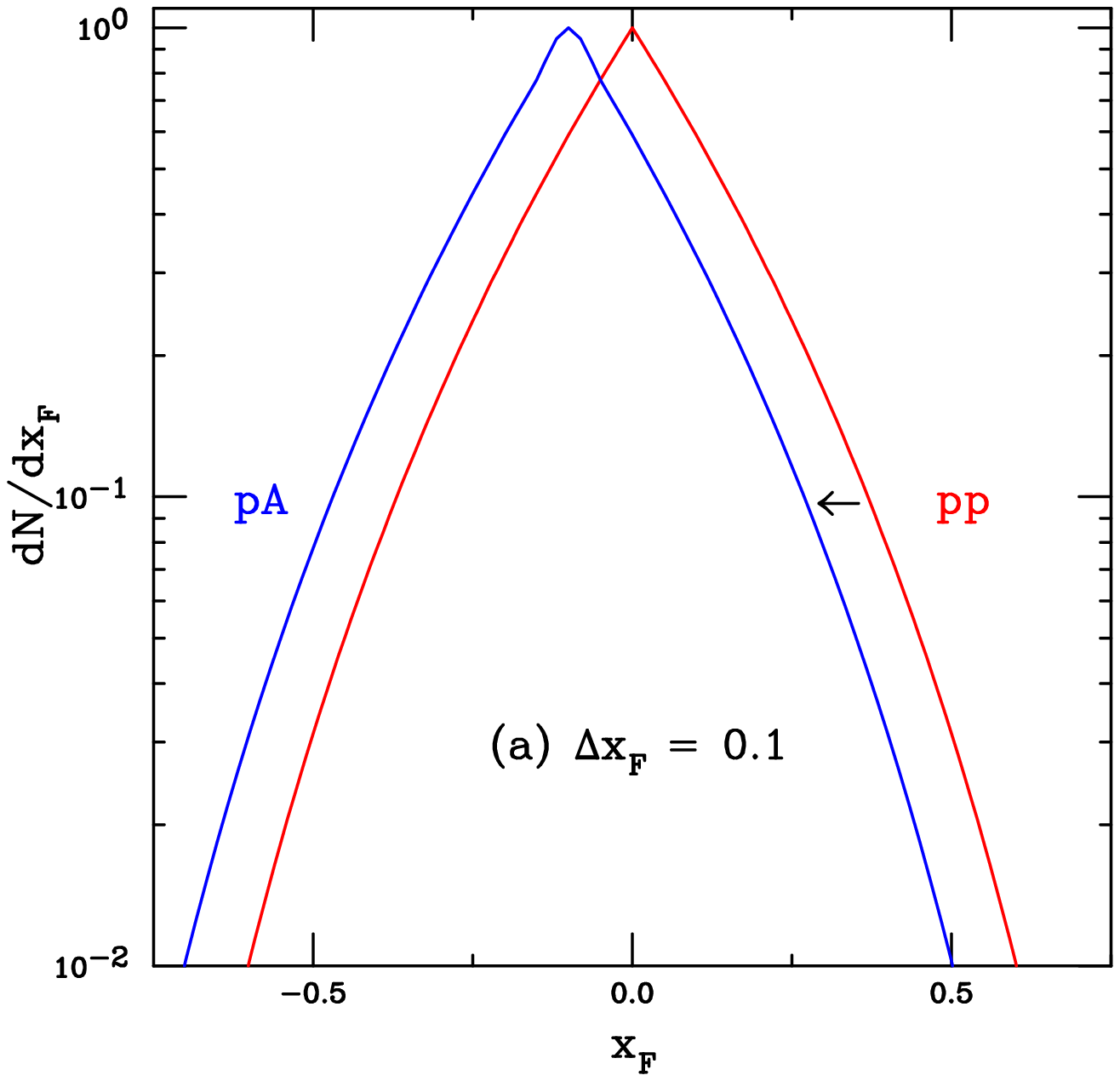}
\includegraphics[width=0.425\textwidth]{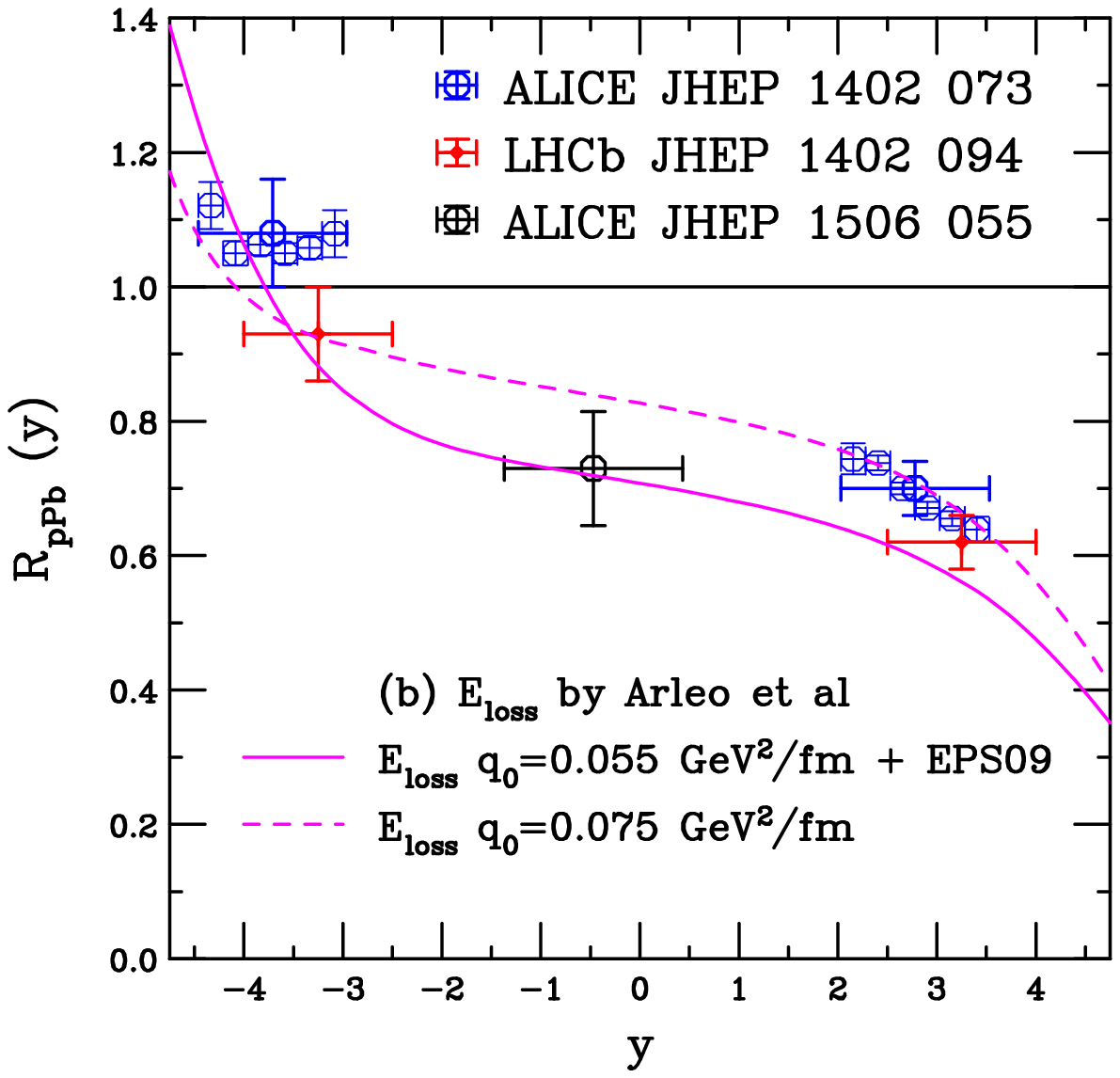}
\caption{(Color online)
(a) Schematic shift in $x_F$ distribution caused by energy loss.
(b) The LHC $J/\psi$ $R_{p {\rm Pb}}(y)$ data from ALICE 
\protect\cite{ALICEpPbpsi,ALICEpPbpsi_pT} and LHCb \protect\cite{LHCbpPbpsi}
compared to energy loss model of Arleo {\it et al.} 
\protect\cite{Arleo:2012rs,Arleo:2013zua}.}
\label{fig:eloss}
\end{figure}
%%%%%%%%%%%%%%%%%%%%%%%%%%%%%%%%%%%%%%%%%%%%%%%%%%%%%%%%%%%%%%%%%%%%%%%%%

Initial-state energy loss in the medium can be
connected to transverse momentum, $k_T$, broadening
in nuclei relative to $pp$ collisions, also known as the Cronin effect 
\cite{Kluberg:1977bm}.

Final-state nuclear absorption, which affects only quarkonium states, involves 
CNM breakup of the (proto)quarkonium state.  Absorption is related to the size
and production mechanism of the interacting state and can be described by a
survival probability, $S_A^{\rm abs} = \exp\{ -\int_z^\infty dz' \rho_A(b,z')
\sigma^C_{\rm abs}(z-z')\}$  where $z'$ is the production point and $z$ is the
dissociation point; $\rho_A(b,z')$ is the nuclear matter density; and 
$\sigma^C_{\rm abs}$ is the effective absorption cross section for quarkonium
state $C$ \cite{Vogt:2001ky}.  The $J/\psi$
has been most studied.  Larger effects, at least at midrapidity, have been
seen for the $\psi'$ \cite{Leitch:1999ea,Adare:2013ezl}.  
The $A$ dependence of the $\chi_c$ is still largely unknown.  The PHENIX
$\chi_c$ result \cite{Adare:2013ezl} has large uncertainties.

%%%%%%%%%%%%%%%%%%%%%%%%%%%%%%%%%%%%%%%%%%%%%%%%%%%%%%%%%%%%%%%%%%%%%%%%%
%\begin{figure}[htb]
%\begin{minipage}[t][7.6cm]{0.5\linewidth}
%\subfloat{\includegraphics[width=5.53cm,height=8cm]{rT_dep_mods.ps}}
%\end{minipage}%
%\begin{minipage}[b][7.6cm]{0.5\linewidth}
%\subfloat{\includegraphics[width=5.53cm,height=4cm]{y_mods_sigma_simple.ps}}
%\vfill
%\subfloat{\includegraphics[width=5.53cm,height=4cm]{sigAbs_yCMS_EPS09.eps}}
%\end{minipage}%
%\caption{(Color online) The gluon modification factor $R_g(r_T)$ from the best fit global $R$ and $d$ (dark solid line) from Eq.~(\ref{eqn:modshad_sharp}), along with results for all combinations of $R$ and $d$ within the $\Delta\chi^2=2.3$ fit contour(thin histograms).  The modification with $T_A^n(r_T)$ ($n=15$), Eq.~(\ref{eqn:modshad_power}), is shown by the light solid line. The  dashed line is the EPS09s impact parameter dependence. From Ref.~\protect\cite{McGlinchey:2012bp}. Comparison of $\sigma_{\rm abs}(y)$ extracted from the PHENIX data assuming a linear dependence on nuclear thickness, Eq.~(\ref{eqn:modshad_power}), and using global values of $R$ and $d$ in Eq.~(\ref{eqn:modshad_sharp}). From Ref.~\cite{McGlinchey:2012bp}. (Right)  Dependence of $\sigma_{\rm abs}^{J/\psi}$ on $y_{\rm cms}$ for all available fixed-target data sets including EPS09 shadowing.  The  asymmetric Gaussian shape of the curves is fixed by the E866 and HERA-B data.  From Ref.~\cite{Lourenco:2008sk}.}
%\label{fig:abs}
%\end{figure}
%%%%%%%%%%%%%%%%%%%%%%%%%%%%%%%%%%%%%%%%%%%%%%%%%%%%%%%%%%%%%%%%%%%%%%%%%

Previous studies have shown the absorption cross section to depend on
rapidity (or $x_F$).  Increased effective absorption at backward rapidity 
may be due to interaction or conversion inside the target while increased 
effective absorption at forward rapidity may be due to energy loss.   The 
effective $J/\psi$ absorption cross section obtained from the PHENIX data, in
combination with the EPS09 shadowing parameterization, exhibits a somewhat
parabolic shape, with a minimum at midrapidity and increasing at forward and
backward rapidity, according to the intuition above.  This result was found
to be independent of the chosen centrality dependence \cite{McGlinchey:2012bp}.
%with the gluon centrality
%dependence of Fig.~\ref{rtmods} is shown as a function of rapidity on the 
%left-hand side of Fig.~\ref{psiabsfig}, both for a linear dependence on 
%$T_A(r_T)$, $n=1$ in Eq.~(\ref{eqn:modshad_power}), and the core 
%parameterization of Eq.~(\ref{eqn:modshad_sharp}).  A stronger $y$ dependence
%and thus a larger effective $\sigma_{\rm abs}$ is seen both at forward and 
%backward rapidity.  While there are some differences, results from 
%the two shadowing 
%parameterizations are in agreement with each other.  
%The dependence of $\sigma_{\rm abs}^{J/\psi}$ on rapidity at different values of $\sqrt{s_{_{NN}}}$ is shown on the right-hand side of Fig.~\ref{psiabsfig}.  In this calculation, 
This result is in agreement with the energy dependence obtained at midrapidity 
in Ref.~\cite{Lourenco:2008sk}, the best fit $\sigma_{\rm abs}^{J/\psi}$ 
decreases with increasing $\sqrt{s_{_{NN}}}$.  In the same paper, the increase
of the effective absorption cross section for $y>0$, turns
on closer to $y=0$ for lower $\sqrt{s_{_{NN}}}$, in addition to the
stronger absorption at lower energies \cite{Lourenco:2008sk}.  

Intrinsic charm is still controversial but could play a role in the dependence
of forward $J/\psi$ production \cite{RVE866,JPL}.  See Ref.~\cite{IC_review} 
for a recent review and Ref.~\cite{StanSusan} on the interpretation of global 
fits of the charm structure function.

So far, only charmonium production has been discussed in detail.  Of the
aforementioned effects, only absorption by nucleons and comovers is ineffective
for open charm.  The effect of $k_T$ broadening was applied to $J/\psi$
production to take the place of low $p_T$ resummation on the $c \overline c$
pair $p_T$ distribution in $pp$ collisions.  No modification was assumed in
$pA$ collisions.  The $k_T$ broadening applied to open charm was assumed to be
the same as in $J/\psi$.  In addition, fragmentation must also be taken into
account.  Previously $k_T$ broadening and Peterson fragmentation were employed
to explain the $p_T$ dependence of $D$ meson production in $pA$ and
$\pi A$ interactions \cite{MNR2}.  The strong momentum shift of the Peterson
parameter, $\epsilon = 0.06$, required a relatively large $k_T^2$ to counteract
the fragmentation and bring the result into agreement with the data, indeed
making the combined effect similar to the original charm quark distribution
itself.  More recently, the fragmentation function employed in the FONLL
approach, without any $k_T$ broadening, was shown to be softer \cite{CNV}.
This behavior can be reproduced in the MNR code \cite{MNR} with a reduced
Peterson parameter, $\epsilon = 0.008$.

\begin{figure}[htb]
\centering
\includegraphics[width=0.425\textwidth]{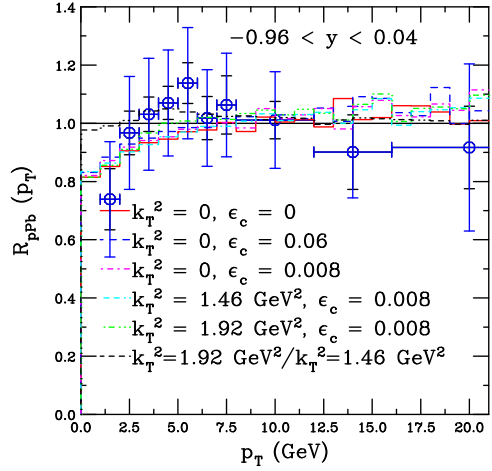}
\caption{(Color online)
The ratio $R_{p{\rm Pb}}(p_T)$ for $\sqrt{s_{NN}} = 5$ TeV 
with EPS09 NLO shadowing only
(red), $k_T^2 = 0$ and $\epsilon_c = 0.06$ (blue), $k_T^2 = 0$ and 
$\epsilon_c = 0.008$ (magenta), $k_T^2 = 1.46$ GeV$^2$ and $\epsilon_c = 0.008$ 
(cyan) and $k_T^2 = 1.92$ and $\epsilon_c = 0.008$ (green).  The last
result (black) assumes a larger intrinsic $k_T$ kick in $p+$Pb than in $pp$.
The ALICE results for average $D$ mesons \protect\cite{ALICEpPbD} is also shown.
%The statistical errors are shown in X, the systematic uncertainties are shown
%in Y.
}
\label{fig:opencharm} 
\end{figure}

At collider energies one might expect that the effect of $k_T$ broadening is
diminished. While this is certainly the case at high $p_T$, in the low $p_T$
domain of the ALICE measurements, it is still important.  To investigate the 
effects of fragmentation and $k_T$ broadening on the nuclear modification 
factor, in Fig.~\ref{fig:opencharm}, we show the calculated $p+$Pb to $pp$ 
ratios compared
to the ALICE data \cite{ALICEpPbD}, averaged over $D^0$, $D^+$ and $D^*$ and
their antiparticles.  The same conditions are used for both $p+$Pb and $pp$
collisions (aside from the EPS09 central set employed in $p+$Pb) for all 
histograms except the black dotted histogram where a large $k_T^2$, as might be
expected, is used in $p+$Pb collisions.  This result shows that, as long as
other effects are kept constant in vacuum and in media, shadowing
dominates and all the values of $R_{p{\rm Pb}}(p_T)$ are consistent.  However,
assuming $k_T$ broadening increases in medium counters the shadowing effect
in the ratio.  The large uncertainties in the data are consistent with all
the calculations, including the one with enhanced broadening.

In conclusion, numerous cold nuclear matter effects have been postulated.
The most studied, nuclear shadowing, is important, but its strength is unclear.
Fixed-target results suggest other effects, including energy loss or intrinsic
charm, could be more important in the forward region.  Clear evidence for
intrinsic charm requires more data focused in the $x$ and $Q^2$ region where it
should dominate.  The effects of energy loss in cold matter described here
should be verified with other processes.  A new global analysis of all data,
including those of the LHC, should be done to provide a new shadowing benchmark.

%%%%%%%%%%%%%%%%%%%%%%%%%%%%%%%%%%%%%%%%%%%%%%%%%%%%%%%%%%%%%%%%%%%%%%%%%%%
%%%%%%%%%%%%%%%%%%%%%%%%%%%%%%%%%%%%%%%%%%%%%%%%%%%%%%%%%%%%%%%%%%%%%%%%%

%%%%%%%%%%%%%%%%%%%%%%%%%%%%%%%%%%
%\Acknowledgements

%This work was performed under the auspices of the U.S.\
%Department of Energy by Lawrence Livermore National Laboratory under
%Contract DE-AC52-07NA27344 and supported by the U.S. Department of Energy, 
%Office of Science, Office of Nuclear Physics (Nuclear Theory) under contract 
%number DE-SC-0004014.

%%%%%%%%%%%%%%%%%%%%%%%%%%%%%%%%%%

\end{document}

%% file: econfmacros.tex
\def\beq{\begin{equation}}
\def\eeq#1{\label{#1}\end{equation}}
\def\eeqn{\end{equation}}

\def\beqa{\begin{eqnarray}}
\def\eeqa#1{\label{#1}\end{eqnarray}}
\def\eeqan{\end{eqnarray}}

\let\bar=\overbar

\def\Dslash{\not{\hbox{\kern-4pt $D$}}}
\def\dslash{\not{\hbox{\kern-2pt $\del$}}}

\def\msb{{\bar{\ssstyle M \kern -1pt S}}}